\documentclass{raa}
\usepackage{graphicx,times}

\begin{document}

\title{LAMOST Experiment for Galactic Understanding and Exploration (LEGUE)}\subtitle{The survey science plan}
\volnopage{Vol.0 (2012) No.0,000--000}
\setcounter{page}{1}

\author{Licai Deng\inst{1} 
\and Heidi Jo Newberg\inst{2} 
\and Chao Liu\inst{1} 
\and Jeffrey~L.~Carlin\inst{2}
\and Timothy~C.~Beers\inst{3} 
\and Li Chen\inst{4} 
\and Yuqin Chen\inst{1}
\and Norbert Christlieb\inst{5}
\and Carl J. Grillmair\inst{6} 
\and Puragra Guhathakurta\inst{7}
\and Zhanwen Han \inst{8} 
\and Jinliang Hou\inst{3} 
\and Hsu-Tai Lee\inst{9}
\and S\'ebastien L\'epine\inst{10} 
\and Jing Li\inst{1,11}
\and Xiaowei Liu\inst{12} 
\and  Kaike Pan\inst{13} 
\and J. A. Sellwood\inst{14} 
\and Hongchi Wang \inst{15} 
\and Fan Yang\inst{1}
\and Brian Yanny\inst{16} 
\and Haotong Zhang\inst{1}
\and Yueyang Zhang\inst{1} 
\and Zheng Zheng\inst{17} 
\and Zi Zhu\inst{18}
}

\institute{Key Lab for Optical Astronomy, National Astronomical Observatories, Chinese Academy of Sciences (NAOC) email: licai@bao.ac.cn \\ \and
Department of Physics, Applied Physics, and Astronomy, Rensselaer Polytechnic Institute, 110 8th Street, Troy, NY 12180, USA\\ \and
National Optical Astronomy Observatory, Tucson, AZ 85719, USA\\ \and
Shanghai Astronomical Observatory, Chinese Academy of Sciences, 80 Nandan Road, Shanghai 200030, China\\ \and
University of Heidelberg, Landessternwarte, Kšnigstuhl 12, D-69117 Heidelberg, Germany\\ \and
Spitzer Science Center, 1200 E. California Blvd., Pasadena, CA 91125, USA\\ \and
Department of Astronomy and Astrophysics, University of California Santa Cruz, 1156 High Street, Santa Cruz, CA 95064, USA \\ \and
Yunnan Astronomical Observatory, Chinese Academy of Sciences, Kunming 650011, China\\ \and
Academia Sinica Institute of Astronomy and Astrophysics, Taipei 106, China\\ \and
American Museum of Natural History, Central Park West at 79th Street, New York, NY 10024, USA\\ \and
School of Physics and Electronic information, China West Normal University, 1 ShiDa Road, Nanchong, Sichuan 637000, China\\ \and
Department of Astronomy \& Kavli Institute of Astronomy and Astrophysics, Peking University, Beijing 100871, China\\ \and
Apache Point Observatory, PO Box 59, Sunspot, NM 88349, USA.\\ \and 
Department of Physics and Astronomy, Rutgers University, 136 Frelinghuysen Road, Piscataway, NJ 08854-8019, USA\\ \and
Purple Mountain Observatory, Chinese Academy of Sciences, Nanjing, Jiangsu 210008, China\\ \and
Fermi National Accelerator Laboratory, Batavia, IL 60510, USA\\ \and
Department of Physics and Astronomy, University of Utah, Salt Lake City, UT 84112, USA\\ \and
School of Astronomy and Space Science, Nanjing University, Nanjing 210008, China
}


\abstract{
We describe the current plans for a spectroscopic survey of millions of stars in the
Milky Way galaxy using the Guo Shou Jing Telescope (GSJT, formerly the
Large Area Multi-Object Spectroscopic Telescope - LAMOST).  The survey
will obtain spectra for 2.5 million stars brighter than $r<19$
during dark/grey time, and 5 million stars brighter than $r<17$ or
$J<16$ on nights that are moonlit or have low transparency.  The
survey will begin in fall of 2012, and will run for at least four
years. The telescope design constrains the optimal declination range for observations to
$10^\circ<\delta<50^\circ$, and site conditions lead to an emphasis on stars in the direction of
the Galactic anticenter.  The survey is divided into three parts with
different target selection strategies: disk, anticenter, and spheroid.
The resulting dataset will be used to study the merger history of the
Milky Way, the substructure and evolution of the disks, the nature of
the first generation of stars through identification of the lowest
metallicity stars, and star formation through study of open clusters
and the OB associations.  Detailed design of the LEGUE survey will be
completed after a review of the results of the pilot survey in summer
2012.}

\authorrunning{Licai Deng, Newberg et al.}
\titlerunning{LEGUE science}

\maketitle

\section{Introduction}

Study of stars in the Milky Way galaxy is critical to understanding how galaxies
form and evolve.  Through study of galaxy formation, we test models of dark matter,
gravitational collapse, hydrodynamics of the gas, stellar formation and feedback
(including properties of the first generation of stars and enrichment of the
interstellar medium through supernova explosions).  The Milky Way is the only
galaxy we can study in enough detail that these models can be tested in six phase-space
dimensions.  Only recently have large photometric sky
surveys, including
the Sloan Digital Sky Survey (SDSS; York et al. 2000, Gunn et al. 2006) and the 2~Micron All-Sky Survey (2MASS; Skrutskie et al. 2006), made it possible to
piece together the structure of the Milky Way star by star.  The Sloan Extension for
Galactic Understanding and Exploration (SEGUE; Yanny et al.~2009) produced sparse samples of stellar spectroscopy.  Large spectroscopic
surveys like the RAdial Velocity Experiment (RAVE; Steinmetz 2003), which targets only the brightest
stars; the APO Galactic Evolution Experiment
(APOGEE,
Allende Prieto et al. 2008), which is observing late-type giant stars in the infrared; and the High Efficiency Resolution Multi-Element Spectrograph (HERMES; Freeman \& Bland-Hawthorn 2008)
for the AAT, which will deliver high accuracy metallicity measurements for a million bright ($V<14$) Galactic stars, are currently in progress. With the holistic models of the Galaxy available (e.g. Sharma et al. 2011), it is possible to restrict artifacts that usually happen  the surveys  of the Galaxy (including LEGUE) when a limited number of stars will be sampled. The Gaia satellite (Perryman et al. 2001; Katz et al. 2004) will measure useful proper
motions for one billion
Galactic stars to $\sim20$th magnitude, radial velocities of $\sim150$~million stars to $\sim16$th mag, and stellar atmospheric parameters for $\sim5$~million stars to $\sim12$th mag. However, there is an compelling case for much larger, deeper, and denser spectroscopic
surveys of Milky Way stars. 

We plan to study the structure of the Galactic halo (both the smooth component of
the spheroid and the lumpy sub-structures) and disk components (including
star-forming regions and open clusters).  The revealed structure will
inform our models of star formation, the formation history of the Galaxy, and the
structure of the gravitational potential, including the central black hole and
(sub)structure of the dark matter component.

In this work we discuss the LAMOST Experiment for Galactic Understanding and Exploration (LEGUE).
The four year Galactic structure survey plan presented here includes
spectra for
2.5 million stars brighter than $r<19$ and an additional 5 million stars brighter than $r<17$; the actual distribution of magnitudes observed will depend on the throughput of the GSJT system (Zhao et al. 2012, this volume). Most of the
stars in the survey will be observed at $R= \lambda/\Delta \lambda = 1800$ which is achieved by placing a 2/3 width slit in front of the fibers; the grating is $R=1000$.
Additional $R=5000$ gratings will be obtained part way through
the survey and will be used particularly at low Galactic latitude where the star density is higher.

The LEGUE survey is divided into three parts: the spheroid, the disk, and the anticenter.  The spheroid survey covers $|b|>20^\circ$; the anticenter
survey covers Galactic latitude $|b| \le 30^\circ$, and longitude
$150^\circ \le l \le 210^\circ$; and an extended disk survey
covers as much of the low latitude sky ($|b| \le 20^\circ$) as is
available from Xinglong Station; the $20^\circ \le l \le 80^\circ$
region of the disk will be poorly sampled due to a limited number of
clear nights in summer (see Yao et al. 2012 for more on site conditions).  Each of these portions of the survey have somewhat different target selection algorithms, which will be similar to the target selection algorithms implemented in the pilot survey, and described in companion papers (Carlin et al. 2012, Yang et al. 2012, Zhang et al. 2012, Chen et al. 2012).

\section{Science Goals}

With the data for a huge number of stars from
the LAMOST spectroscopic survey, stellar kinematics can be
calculated and the metallicity distribution function (MDF) in the
Galaxy will be obtained. This will allow us to systematically
investigate the space density, Galactocentric rotation velocity and
velocity ellipsoid, and chemical abundance of stars as a function of
position in the Galaxy. These provide important constraints on the
present models of the Galactic structure, formation history,
kinematical and dynamical evolution, chemical evolution, and the
dark matter distribution in the Milky Way.

The primary science drivers of the LAMOST Galactic structure survey are:

\begin{enumerate}

\item Search for extremely metal poor stars in the Galactic spheroid

\item Kinematic features and chemical abundances of the thin/thick disk stars, with the goal of deriving the mass distribution (including the dark matter mass), the dynamical and chemical evolution, the structure and the origin of the Galactic disks.

\item A thorough analysis of the disk/spheroid interface near the Galactic anticenter, with the
goal of determining whether previously identified anticenter structures are tidal debris, or
whether they are part disk structures

\item Discovery of stellar moving groups that may be associated with dwarf galaxies, and follow-up observations of known
streams and substructures in the Galactic spheroid

\item Survey of the properties of Galactic open clusters, including the structure, dynamics and evolution of the disk as probed by open clusters;

\item Search for hypervelocity stars and determination of their creation mechanism

\item Survey the OB stars in the Galaxy, tracing the 3D extinction in the Galactic plane

\item A complete census of young stellar objects across the Galactic Plane, which provides
important clues to studies of large-scale star formation and the history of Galactic
star formation.

\end{enumerate}

A typical image of a randomly chosen field of the Milky Way will contain
stars at many distances from local disk to distant halo, and may also contain groups of stars of a variety of origins.  While these groups will
be well mixed and indistinguishable from multi-color imaging alone, the
addition of kinematic information and spectroscopic stellar atmospheric
parameter information ($\mathrm{[Fe/H]}$, T$_{\tiny\mathrm{eff}}$, log $g$, $[\alpha/$Fe]) make possible
the identification of a common origin for groups of stars.  If one has
a large enough kinematic sample, such as that proposed here for LAMOST, one
can begin to trace the origin and build up of the Galaxy itself
and explore the role that individual bursts of star formation at different times played in the assembly of the thick disk and halo.
This is a study that can only be done with spectra of hundreds of thousands to
millions of stars.

It was a tremendous asset of the SDSS imaging survey that it covered
a large ($>8000\mathrm{deg^2}$) region contiguously, with no
significant holes in area.  For example, this allowed clear unambiguous discovery
of faint long streams tracing around our Galaxy; e.g., the Grillmair
and Dionatos 63-degree stream (Grillmair \& Dionatos 2006), and the
Orphan Stream (Grillmair 2006, Belokurov et al. 2006), among others.
Without a large contiguous picture it would have been difficult if
not impossible to piece together the very low contrast density
enhancements that make up these faint structures.  Since LAMOST can
cover a similar large area spectroscopically, one could look for
kinematic streams in radial velocity (RV hereafter) and position
which are contiguous across the sky, and connect pieces of
structures which it would otherwise not be possible to unambiguously
associate with each other.

The SEGUE survey was designed in an era when very large density substructures were
beginning to be discovered in the Milky Way's stellar halo.  It was designed to sample
these structures on the largest scales, and therefore consisted of many pencil beam
surveys that covered a large range of distances.  The stars to be observed were chosen
using more than a dozen separate and possibly overlapping highly complex target selection
categories, each optimized for a different science goal.  While these choices were
reasonable at the time, we now need a large density of spectra in each volume of the
Galaxy to study lower contrast substructure and discover substructure in the Galactic disk.
The much larger sample of stars in the LEGUE survey will therefore be selected with more
contiguous sky coverage and with higher density spatial sampling.

Although the LAMOST survey will be more complete than SEGUE,
three factors keep us from a complete survey in magnitude and sky area
covered: (1) LAMOST fibers do not move more than 3.15' from their nominal positions, (2) we
will only observe 2\% of the estimated $5\times 10^8$ stars in the available $\sim 2 \pi$
steradians, and (3) fields are circular, and the centers are constrained to be centered on a bright star that can be used for real-time active mirror corrections.  Our survey will
not be complete in sky coverage.  However the selection algorithm will be a simple
function of color and magnitude, with weighted random sampling.

Below we highlight several of the interesting science cases that will be addressed with LEGUE spectroscopy.

\subsection{Metal-poor stars}

Metal-poor stars provide the fossil record of the creation and evolution
of the elements from the earliest times, and thus play an important role
in the study of early Galactic chemical evolution. In particular, the
most metal-deficient and hence oldest stars in the Galaxy provide
information about their Population III progenitors, which must have
existed in the past but, if there were no low-mass counterparts, do not
exist anymore today. Due to the rapid decline of the metallicity
distribution function (MDF) of the Galactic halo towards low
metallicities, extensive, wide-field objective-prism surveys such as the
HK survey (e.g., Beers et al. 1985,1992) and the Hamburg/ESO Survey
(HES; Christlieb et al. 2008) have been used to identify the most
metal-poor stars. The combined sample of very metal-poor stars (VMP;
[Fe/H] $< -2.0$) discovered by these efforts to date now exceeds 3000.
Most recently, the massive spectroscopic effort from SDSS, and in particular SEGUE,
have increased the number of known VMP stars to over 30000, including
on the order of 1000 extremely metal-poor stars (EMP; [Fe/H] $< -3.0$),
and a least a handful of stars likely to be even lower metallicity.

High-resolution spectroscopic studies of subsets of these have produced
detailed studies of the nucleosynthesis process in the early Galaxy and
the Big Bang (see Beers \& Christlieb 2005; Frebel \& Norris 2011). For
example, abundances of $^7$Li for metal-poor stars provide constraints on
Big Bang nucleosynthesis (e.g., Asplund et al. 2006; Mucciarelli et al.
2012), and the nature of the slow (s-) and rapid (r-) neutron-capture
processes can be investigated by metal-poor stars enhanced in the
respective elements (e.g., Sneden et al. 2008).

Extrapolating from the number of EMP, Ultra Metal-Poor (UMP; [Fe/H] $<
-4.0$) and Hyper Metal-Poor (HMP; [Fe/H] $< -5.0$) stars found by the
HK, HES, and SDSS efforts, LEGUE is expected to yield 10000 or more
stars with [Fe/H] $< -3.0$, hundreds of stars with [Fe/H] $< -4.0$, and
$\sim$20-30 stars with [Fe/H] $< -5.0$, increasing the number of known stars
in these metallicity ranges by over an order of magnitude.

We will also determine the shape of the low-metallicity tail of the halo
MDF with much higher accuracy than is possible with the limited samples
available today. The MDF provides important information on the formation
and chemical enrichment history of the Galactic halo (e.g., Norris et
al. 2012).

\subsection{The Disk in and beyond the Solar Neighborhood}

We propose to study the kinematic features and the chemical abundances of the Galactic thin/thick disks in the Solar neighborhood and beyond.

\subsubsection{The local dark matter density}

The local mass density of the Milky Way is still an open question since Oort's (1932, 1960) first
attempt to solve the Poisson-Boltzmann equation using the movement of a group of stars near the Sun.
Bahcall (1984), Kuijken \& Gilmore (1989), Holmberg \& Flynn (2000, 2004) etc. also made estimates of
the local matter density using various observed data and methods. Kuijken \& Gilmore (1989), using
well observed radial velocities from a volume-selected K dwarf star sample near the South
Galactic pole, found little evidence of local dark matter. Holmerg \& Flynn (2000, 2004) used
Hipparcos data, which have distance measurements but are not guaranteed to be a complete sample, and found
a significant amount of dark matter in the Solar neighborhood. Recently, Garbari et al.
(2012) and Zhang et al. (2012) revisit this question with
improved methods but still do not resolve the issue.

In a new attempt to estimate the local mass density, we propose to observe about 100,000 nearby
bright stars, selected from Tycho-2 catalog (H\o g et al.\ 2000), with accurate proper motions, and
additionally a complete set of dwarf stars toward the North Galactic pole with LAMOST. The metallicity
corrected photometric distance and the vertical velocity dispersion as a function of the distance to the
disk mid-plane will be obtained. The spectra of these stars must have S/N and spectral
resolution sufficient to distinguish giants from dwarfs and to obtain radial velocities to a precision of
better than 5 km s$^{-1}$, which is possible with the LEGUE $R=$5000 gratings. With the vertical velocity,
which is contributed by the radial velocities
in the North Galactic pole, and by the combination of the radial velocity, distance and the
proper motion in other directions one can improve the local disk matter density measurement with
a larger and more complete sample.

\subsubsection{Local velocity substructure in the disk}

These data also enlarge the sample of the nearby bright stars and will be a significant improvement
over the Geneva-Copenhagen Survey (Nordstr\"om et al.\ 2004, hereafter GCS). The $\sim 14,000$
F \& G dwarfs in the GCS survey have full phase-space coordinates; a follow-up survey by
Famaey et al.\ (2005) has similar data for $\sim 6000$ K \& M stars. As a result of these studies,
there are very few stars with HIPPARCOS distances that do not have well-determined space motions.  
These samples have revealed intriguing structure in the distribution of their Galactic U \& V
velocities, that contains important information about the dynamical history of the Milky Way through
resonance scattering (Dehnen 2000, Helmi et al.\ 2006). For example, Sellwood (2010) has been able
to identify strong evidence for a recent Lindblad resonance in the solar neighborhood. And Antoja et al. (2012) reveals the similar substructure beyond Solar neighborhood with RAVE data. 

The phase-space information of the proposed LAMOST nearby bright star survey can be converted
to action-angle variables to reveal finer details of the
phase space structure that will enable the dynamical origin of the
features to be identified.  Hahn et al.\ (2011) have already performed
this analysis on a sample of $< 7000$ main sequence stars withn
200~pc of the Sun from the RAVE (DR2) survey and the M-dwarf catalog
derived from SDSS by West et al. (2011).  They found evidence
confirming the structure already identified in Sellwood (2010), but
many more stars are needed to in order to find less prominent
substructures.  The on-going RAVE survey will substantially increase
the sample from the southern sky and will nicely complement the LAMOST
northern sky survey. It is worthy to note that searching for substructure in phase-space depends heavily on the survey selection functions, which have been carefully designed for this kind of science (Carlin et al. 2012).

\subsubsection{Disk structure from red clump stars}

Apart from the nearby stars, there are also a large number of intrinsically luminous giant stars in the disk
that can be observed by LAMOST. They bring kinematic information a few kpc beyond the Solar neighborhood.
Some of the luminous stars, particularly the red clump stars, are good distance indicators.  We expect
to observe about 500,000 red clump stars with $V<17$ in the anticenter region:
$|b|<20^\circ$ and $90^\circ<l<220^\circ$.
The red clump stars at low Galactic latitude can be easily selected from the 2MASS
color-magnitude diagram (L\'opez-Corredoira et al. 2002). Nearby dwarf stars can overlap in intrinsic
color with the red clump stars, but as the latter get reddened by extinction in the mid-plane,
the dwarf stars can be easily removed. This color selection of red clump stars is thus very
efficient with only little contamination by other giant stars and remaining dwarf stars (Liu et al. 2012).

Similar to the substructure found in the U-V distribution in the Solar neighborhood, the resonance
of the Galactic bar and spiral arms may also induce substructure in the velocity distribution
in the outer disk. Moreover, the velocity distribution of the stars right on the spiral arms
will also show some special substructure, e.g. arcs, due to the correlation of the radius and
the orientation of the stellar orbits (Quillen et al. 2011). Indeed, Liu et al. (2012)
found evidence of the resonance in Galactocentric radius of 10--11\,kpc in the anti-center
direction from the distribution of the radial velocity of only $\sim800$ red clump stars observed by
MMT/Hectospec. Since LAMOST can extend the disk observations to more than 100$^\circ$ in
azimuth, it will provide clearer evidence of the resonance.  Combining these with
the velocity distribution in the Solar neighborhood derived from LAMOST bright spectra, one can
model the rotation pattern speeds, the strength and the current phase of the Galactic bars and
the spiral arms.

The red clump stars will cover about 4\,kpc in Galactocentric radius, from 8\,kpc (Solar radius) to
12\,kpc. The proper motions have been provided for most of them in PPMXL catalog (R\"oser et al. 2010).
Once red clump stars are reliably identified along the line of sight to the Galactic anticenter from LAMOST spectra, their proper motions can be used to derive the Galactic rotation curve. The Galactic rotation curve has been
measured from HI, CO and masers (e.g. Sofue et al. 2009, Sakai et al. 2012). However, because
HI and CO suffer high distance uncertainty, the rotation curve derived from
these objects is not reliable.  Though the distance to the masers can be
accurately measured from trigonometric parallax, only very few masers have been measured this way;
furthermore, since most of them are located on spiral arms, they are not the best tracers to continuously cover all Galactocentric radii. We propose to measure the rotation curve
using the radial velocities of the red clump stars in the disk. Since the distance can be well
determined, the rotation curve derived from this sample is expected to be significantly more accurate.
The improved rotation curve will make it possible to decompose the disk and
halo components of the Galactic gravitational potential. The shape and mass of the dark
matter halo, including the local dark matter, will be estimated. In the light of
Rix \& Zaritsky's (1995) work on external galaxies, it is interesting to ask whether the
disk of our Galaxy is also lopsided. This could be studied with the kinematic features of the
disk stars in the LAMOST survey data combined with other survey catalogs, e.g.
APOGEE (Allende Prieto et al. 2008).

\subsubsection{Radial migration}

Though the thick disk was discovered by Gilmore \& Reid in 1983, its origin is still a puzzle. There
have been dozens of models of thick disk formation (see the annual review by
Majewski 1994).  Today, some of them are out of date and four main scenarios are thought to
be the most likely: 1) accreted debris of the disrupted satellites (Statler 1988; Abadi et al. 2003);
2) heating of the pre-existing thin disk by the merging satellites (e.g. Quinn,
Hernquist \& Fullagar 1993; Villalobos \& Helmi 2008); 3) in-situ triggered star formation during and
after a gas-rich merger (e.g. Jones \& Wyse 1983; Brook et al. 2004); and 4) in-situ formation
of the thick disk through radial migration of stars as a consequence
of corotation resonance with transient spiral structures (Sellwood
\& Binney 2002), bar structures (Minchev \& Famaey 2010), or orbiting
satellites (Quillen et al. 2009).  Observational evidence of radial migration has been reported by
Yu et al.\ (2012), Lee et al.\ (2011), Loebman et al.\ (2011), Liu \& Van de Ven (2012) etc.
Moreover, Liu \& Van de Ven (2012) found that radial migration is not the only channel to
form the thick disk; the stars on eccentric orbits with [Fe/H]$<-0.5$\,dex may have originated in a gas-rich merger.

The chemical abundance of the stars can be considered the ``real" integral of motion, since it does not
change over most of the lifetime of a star. Abundances tell us about the place in which
the star formed. The chemo-kinematic study on both the thin and thick disk stars can be conducted based on
the data from the LAMOST spectroscopic survey.  This, for instance, is the best way to investigate
radial migration from observations. Sch\"onrich \& Binney (2009) have set up a chemical evolution model
taking into account radial migration. Loebman et al. (2011) pointed out that the $\alpha$-abundance
is a valuable indicator of age that can be used to help to confirm the radial migration.
Sales et al. (2009), Dierckx et al. (2010), and Liu \& Van de Ven (2012) indicate that eccentricity
could also be a powerful
kinematic feature for detecting radial migration. LAMOST survey data in the disk will provide sufficient
samples covering not only the Solar neighborhood but also a few kpc beyond. The data will be used to
investigate the variation
of the abundance features, e.g. metallicity distribution function, with spatial and kinematic features
and subsequently build the chemo-dynamical evolution model of the disk.

Besides providing valuable data to the international astronomical community, this survey, combined with
the future GAIA data, gives a powerful way to describe the spatial positions and other properties
for stars with different abundances. These provide important constraints on the present models of
the disk structure, formation history, and chemo-dynamical evolution of the Milky Way.

\subsubsection{Low-mass stars and the local stellar populations}

All the Galactic stellar populations are dominated by low-mass
stars in the M dwarf/subdwarf range $0.08 M_{\odot} < M < 0.4
M_{\odot}$. By number, they constitute $>70$\% of H-burning stars in
the Galactic disk, and probably a larger, though as yet undetermined,
fraction of H-burning stars in the Galactic halo. M dwarfs/subdwarfs
have long evolutionary timescales and are also fully convective. As a
result, their atmospheric composition is essentially identical to
their primordial chemical makeup. This makes M dwarfs/subdwarfs true
fossils of star formation and chemical evolution in the Galaxy.
M dwarfs have complex spectra dominated by molecular bands of TiO and
CaH which have proven difficult to model. However, recent advances in
optical and infrared spectral analysis have revealed patterns which now
make it possible to evaluate metallicities (Lepine et al. 2007;
Woolfe, Lepine, \& Wallerstein 2009; Rojas-Ayala et al. 2012) and
identify relatively young
($<500$Myr) M dwarfs (West et al. 2008; Schlieder et al. 2012) based
on spectral features in the red-optical and near-IR. The low and
medium-resolution
spectra of M dwarfs/subdwarfs that will be collected in the various
LAMOST surveys will have sufficient resolution and signal-to-noise to
estimate effective temperatures and metallicities, and use gravity and
activity diagnostic features to identify young stars. Preliminary
results from GSJT commissioning data demonstrate that M dwarfs can be
efficiently identified and analyzed (Jing et al., in preparation).

By the nature of this survey, which targets stars at random over an
extended range of colors and magnitudes, large numbers of M
dwarfs/subdwarfs will inevitably be observed in all the fields. Even
in regions dedicated to specific surveys, there will always be a
fraction of fibers that cannot be allocated to primary targets. Those
free fibers will then be placed on available objects following the
general color-magnitude selection scheme, of which a significant
fraction will be low-mass stars. One advantage of M dwarfs/subdwarfs
is their large numbers and relative proximity. Within the magnitude
limits of LEGUE, most observed M dwarfs/subdwarfs will be within 1~kpc
of the Sun, which means that they will generally have relatively
large proper motions (either already recorded or that could be easily
measured), from which full 3D motion can be derived when combined with
the LEGUE radial velocity measurements.

The LEGUE survey will provide a useful complement to the SDSS
survey. Most M dwarfs identified in SDSS lie at high Galactic
latitudes, and thus probe mainly the Galactic old/thick disk (West et
al. 2011). The LEGUE survey will target larger numbers of M
dwarfs/subdwarfs at low Galactic latitudes, and probe deeper into the
young/thin disk. In addition, target selection in SDSS followed a
complicated color-color scheme, and as a result ended up selecting
against the more metal-poor M subdwarfs, whose colors differ
significantly from the colors of metal-rich M dwarfs (Lepine \& Scholz
2008; Lepine et al. in preparation). Indeed the recent claim of
an ``M dwarf problem'' by which the Galactic disk would be deficient in
metal-poor stars (Woolfe \& West 2012), might simply be the result of
the color
selection of SDSS targets. The LEGUE survey would put that conjecture
to the test, and determine the metallicity distribution in the local
disk to unprecedented statistical accuracy.

Combining metallicity data with the full 3D kinematics will provide the
clearest picture of the local and low-mass members of the Galactic
stellar populations, identifying their relative numbers in the Solar
vicinity and mapping their distribution in velocity space. These data
will allow one to search, e.g., for the local signature of Galactic
halo streams (Re Fiorentin et al. 2005, Famaey et al. 2005), for
the kinematic signature of the Galactic spiral arms and bulge
(Sellwood 2011), or signs of accretion events in the disk (Helmi et
al. 2006).




\subsection{Outer disk substructure}

The outer disk is most easily studied in the Galactic anticenter direction, where by virtue of the Sun's position some 8 kpc from the Galactic center, the stars in the outer disk region are relatively nearby. The anticenter direction has proven to be particular rich in substructure, including the Monoceros Ring (Newberg et al. 2002, Yanny et al. 2003, Penarrubia et al. 2005), the Anticenter Stream (Grillmair 2006b, Grillmair, Carlin, \& Majewski 2008, Carlin et al. 2010), the EBS stream (Grillmair 2006b, Grillmair 2011), and the Canis Major Overdensity (CMa, Martin et al. 2004). The Monoceros Ring and CMa have stirred considerable debate, with some investigators claiming that these structures are evidence of warping and/or flaring of the stellar disk (Momany et al. 2006, Natarajan \& Sikivie 2007), while others maintain that they are most likely remnants of merger events (Conn et al. 2007, Butler et al. 2007).


Only with a very large sample of stellar velocity and metallicity measurements in the anticenter direction are we likely to make significant progress in understanding the origins of structures and substructures in the outer disk.
Previously detected substructures are about 10 kpc from the Sun, so the main sequence stars are about 19th magnitude and giant stars are about 15th magnitude, which spans the expected target range of LEGUE.
As the weather at Xinglong Station is statistically most favorable when the anticenter region is overhead, the GSJT survey is particularly well suited for unraveling this complex region and answering many important questions concerning the build-up and evolution of the Galaxy.  

\subsection{Spheroid structure and tidal streams}

Since the discovery of the tidally-disrupting Sagittarius (Sgr) dwarf galaxy in a kinematic
survey of the Galactic bulge (Ibata et al. 1994), photometric surveys
covering large sky areas (e.g., SDSS, 2MASS) have vastly expanded our knowledge of the
spheroid's spatial substructure using carefully
selected stellar tracers.

A spectroscopic survey of large numbers of stars over a contiguous area of
sky is vital to fully characterizing known halo substructures.
While discovery and extensive spatial mapping of halo substructures has been enabled by
photometric surveys such as 2MASS and SDSS, the kinematics of stars in many of the
streams and substructures are poorly known (or completely unknown, in some cases). Using
SDSS/SEGUE as an example, instances where kinematics have been measured for substructures
(e.g., Belokurov et al. 2007, Newberg et al. 2009, Willett et al. 2009, Koposov et al. 2010,
Newberg et al. 2010, Li et al. 2012) have relied on handfuls of SEGUE plates that intersect the
structures. Even if a clear kinematic signature of the stream/substructure can be seen
in the RVs, the sparse spatial coverage makes interpretation of chemodynamical information
difficult.

In addition, a large-area
survey will allow for the {\it discovery} of substructures as kinematical overdensities among
halo stars in a way that sparse sampling cannot. In fact, because phase-space density is
conserved with time (as accreted substructures phase-mix with Milky Way populations), accretion
remnants that are no longer spatially coherent can often be identified from their velocities
(e.g., Helmi \& White 1999). Thus large samples of radial velocities from spectroscopic surveys such
as LAMOST are likely to yield numerous discoveries of accretion relics in the halo (similar to
the ``Cetus Polar Stream" found by Newberg et al. 2009 or the ``ECHOS" discussed in Schlaufman
et al. 2009, both of which came from SEGUE data).  The discovery and mapping of halo substructures over
a large volume of the halo can be used to assess the fraction of the halo that is accreted
and distinguish between expectations from hierarchical structure formation models (e.g.,
Bullock \& Johnston 2005, Cooper et al. 2010, Rashkov et al. 2012).

Tidal streams can be used to constrain the merger history of the Milky Way, and probe the Galactic
dark matter halo.  Measured velocities over
large angular extents of streams can be used in conjunction with distances (and, where available,
proper motions) to fit orbits to streams (see, e.g., Willett et al. 2009, Koposov et al. 2010,
Newberg et al. 2010).  These aid not only aid in recreating the original satellite population, but
can be used as sensitive probes of the shape and strength of the Galactic dark matter halo in
which they are orbiting (e.g., Johnston et al. 1999, Law et al. 2009, Koposov et al. 2010).
The spectra also yield metal abundances -- [Fe/H] is measurable even at low resolution, and abundances
of specific species or families of chemical
elements can be derived at medium-to-high resolution (including alpha-abundances even at SEGUE
resolution of $R\sim2000$; Lee et al. 2011). The chemical signatures of accretion events persist to
some extent in [Fe/H], but can be seen much more clearly in alpha-element abundances
(i.e., [$\alpha$/Fe]), which carry the imprint of the progenitor's enrichment history and can
be used to identify accretion events (or more generally, a global accretion history) through what
has become known as ``chemical tagging" or ``chemical fingerprinting" (e.g.,
Freeman \& Bland-Hawthorn 2002, Johnston et al. 2008, Chou et al. 2010a, b).

The LAMOST/LEGUE survey is particularly well suited to study substructure in the Milky Way
halo, as this survey will take millions of spectra over large contiguous regions in both the
north and south Galactic caps, as well as the low-latitude Galactic anticenter. Because tidal
debris is prevalent at large distances from the Galactic center, it is necessary to obtain
spectra of intrinsically bright stellar tracers, including stars of type A-F (at both the
turnoff and the horizontal branch) and K/M giants. Giant stars are difficult to select
photometrically from among the much larger numbers of dwarfs at similar colors, though the
SDSS $u$-band will aid in the selection of these where it is available. However, giants can
be readily distinguished from dwarfs by surface gravity measurements derived from the LAMOST spectra,
provided that adequate $S/N$ has been achieved to measure reliable stellar parameters. It is
ideal to have sufficient $S/N$ to measure metallicities and surface gravities, but even with
just a radial velocity (which can be measured down to $S/N\sim5$) a lot can be learned about
halo substructures. The LEGUE survey will over-emphasize blue (A, F) stars in its target
selection, as well as observing a large fraction of K/M giant candidates, over a large
contiguous (or nearly contiguous) area of the northern sky, providing a huge dataset to be used
for studies of halo substructure.


\subsection{Open Clusters}

Open clusters (OCs) have long been used to trace the structure and
evolution of the Galactic disk. Open clusters have relatively
large age spans and they can be relatively accurately dated; one
can see OCs to large distances while most of them have average
reddening and distance parameters available; the spatial
distribution and kinematical properties of OCs provide critical
constraints on the overall structure and dynamical evolution of
the Galactic disk; meanwhile, their [M/H] values serve as
excellent tracers of the abundance gradient along the Galactic
disk, as well as many other important disk properties, including the
age-metallicity relation (AMR) and abundance gradient evolution
(Chen et al. 2003).

We expect the LAMOST low Galactic latitude survey to observe $\sim300$
OCs, obtaining stellar radial velocities as well as abundance
information of stars complete to R = 16 mag in the cluster fields.
This will give the largest spectroscopic data set for studying the
Milky Way open clusters.

The large amount of up-to-date homogeneous open cluster data from
LAMOST would lead to the most reliable membership determination for
sample clusters, using accurate radial velocity data. These
will significantly purify the color-magnitude diagrams of hundreds
of open clusters and provide the best basis for obtaining the
essential parameters of clusters, such as distances and ages.

Delicately selected ``standard open clusters" (e.g., M67, see
Smolinski et al. 2011) can play a role as ``models" for calibrating
LAMOST observations and the data-processing pipeline. On the other
hand, well characterized clusters can also be used as calibration
tools when combined with appropriate evolutionary models. Also,
the M67 field could be good targets for detecting or verifying
possible tidal tails of these stellar clusters.

By utilizing radial velocities and proper motions from outside
catalogs, one would be able to trace the structure and kinematics of
the disk as a function of position in the Galaxy, especially in the
following respects: (1) The depth of LAMOST will allow the 3-D
structure of the northern Galactic warp (Guijarro et al. 2010) in
the Galaxy to be traced, using giants and red clump stars in distant
clusters ; (2) Kinematic data of OCs will also allow us to
systematically study the Galactocentric rotation velocity, which
will provide definitive estimates of the mass of the Galaxy, and
also the principal scale parameter $-$ the distance of the Sun from
the Galactic center, $R_\odot$; and (3) By combining chemical
abundances and ages of OCs, one will be able to probe the
correlations between age, velocity, and abundance as a function of
position in the Galaxy (Yu et al. 2012).

\subsection{Hypervelocity Stars}

The existence of hypervelocity stars (HVSs) with velocity higher than the
escape velocity of the Galaxy was predicted by Hills (1988). HVSs
are thought to be ejected from the very center of the Galaxy. The ejection
mechanisms
may include tidal disruption of tight binary stars by the central massive black
hole (MBH) or interaction between single stars and an intermediate mass
black hole inspiralling towards the central MBH (e.g., Hills 1988; Yu 2003).
Since the first discovery of a HVS (SDSS J090745.0+024507, a $g^\prime=19.8$
B9 star at a Galactocentric distance of $\sim$110 kpc with a heliocentric
radial velocity of $\sim$850 ${\rm km\,s^{-1}}$) by Brown et al. (2005), about
19 HVSs
have been found (e.g., Edelmann 2005; Hirsch 2005; Brown et al.2006a,b,
2007a, b, 2012). HVSs, linking the central MBH to the Galactic halo, can be used to probe a variety of properties of the Galaxy
on scales from $\sim5$ pc to $\sim 10^5$ pc (Kenyon et al 2008).
The trajectories of HVSs provide unique constraints on the shape of the
potential of the Galaxy's dark matter halo (Gnedin et al 2005; Yu \& Madau 2007). The
distributions of the velocities and rates of HVSs enable us to pin
down the ejection mechanisms. The density, velocity, and stellar type
distributions of HVSs also tell us about the environment around the
central MBH and the stellar mass distribution near the MBH
(e.g., Brown et al. 2006a; Kollmeier \& Gould 2007; Lu et al. 2007). HVSs may also have connections to
the S stars in the disk near the central MBH (e.g., Lu et al. 2010)
and may provide clues to the growth of the MBH (e.g., Bromley et al. 2012).
A large sample of HVSs is clearly desired for all the above studies.

At present, only one HVS, US 708, is an extremely He-rich sdO star
in the Galactic halo, with a heliocentric radial velocity of
+$708\pm15$\,km s$^{-1}$. Hirsch et al. (2005) speculated that US
708 was formed by the merger of two He WDs in a close binary induced
by the interaction with the SMBH in the GC and then escaped.
Recently, Perets (2009) suggested that US 708 may have been
ejected as a binary from a triple disruption by the SMBH, which
later on evolved and merged to form a sdO star. However, the
evolutionary lifetime of US 708 is not long enough if it originated from
the GC. Wang and Han (2009) found that the surviving companions from
the white dwarf + helium star channel of type Ia supernova (SN Ia) progenitors
have a high spatial velocity ($>$400\,km s$^{-1}$) after a SN Ia explosion,
which could be an alternative origin for HVSs, especially for HVSs
such as US 708 (see also Justham et al. 2009). Considering the
local velocity nearby the Sun ($\sim$220\,km s$^{-1}$), Wang and Han (2009)
found that about 30$\%$ of the surviving companions of SNe Ia may be observed
to have velocity above 700\,km s$^{-1}$. In addition, a SN asymmetric explosion may
also enhance the velocity of the surviving companion. Thus, a
surviving companion star in the white dwarf + helium star channel of SNe Ia may have a
high velocity like US 708.

An alternate origin for HVSs was proposed by Abadi, Navarro, and Steinmetz (2009), who
suggested that some of the stars could be from the tidal disruption of dwarf galaxies.

We estimate the potential of discovering HVSs from a spectroscopic survey of
stars by LAMOST. As a baseline, we assume a survey area of 16,000 square
degrees and a magnitude limit of $r$=19.5. According to the LEGUE halo target
selection (Fan et al. 2012, Carlin et al. 2012), all A- and B-type stars will be
selected and a significant fraction of stars down to M-type will be selected.
Current
searches for HVSs usually preselect B-type stars as targets for spectroscopic
observations to increase the success rate by reducing the contamination of
old halo stars. Identifying hypervelocity stars from the LAMOST survey
is essentially straightforward once the spectra are acquired, and it
is not limited to B-type stars. Based on $\sim$10 HVSs discovered,
Brown et al. (2007b) estimate the space density of 3--4$M_\odot$ (main
sequence B stars, $M_V\sim 0$) HVSs to be (0.077$\pm$0.008)$R^{-2}
{\rm kpc}^{-3}$, where $R$ is the Galactocentric distance in units of kpc.
With a magnitude limit of $r$=19.5, the LAMOST survey can reach $R\sim 80\,$kpc
for B stars, and the above space density implies that $30 \pm 3$ B-type HVSs
are expected to exist in the survey area of 16,000 square degrees.
The LAMOST survey can also
discover a large number of other types of HVSs,
about 180 (830) HVSs down to $0.8M_\odot$ if a Salpeter (Galactic bulge)
initial mass function (IMF) is assumed.
The above predicted numbers are
proportional to the survey
area. They also depend on the limiting magnitude --- one magnitude brighter
would lead to a factor of 1.585 lower in these numbers, since the numbers
are proportional to the Galactocentric distance $R$ that can be reached
given that the number density is proportonal to $1/R^2$.

Compared with the present HVSs sample, the LAMOST survey will likely increase
the HVSs sample by about one order of magnitude or more. In addition to this,
LAMOST will discover a large sample of bound HVSs (with radial velocities in
the range between $\sim$275 and $\sim$450 ${\rm km\, s^{-1}}$; e.g.,
Brown et al. 2007a) and possibly binary HVSs. The large statistical sample
of HVSs of different types discovered by LAMOST will be a powerful tool
to deepen our understanding of the properties of the Galaxy, from the very
center to the outer halo.


\subsection{Three dimensional extinction map across the disk}

Due to the existence of tremendous amounts of dust, it's extremely
difficult, while important, to trace the three dimensional dust
distribution along the disk. Kiszkurno et al (1984) presented a 3D
distribution of extinction in the Galactic disk based on UBV
photometry of a limited number of OB stars. However, both the
spatial coverage and number density of the sample of OB stars is
low, which makes it difficult to perform accurate extinction
corrections.

The magnitude limited spectroscopic survey allows for the first
systematic census of OB stars in the Galactic disk. The high
luminosity of OB stars helps to trace three dimensional extinction
at larger distances, which provides a key data set for subsequent
studies of objects in the Galactic disk, including exploration of the structure and
distribution of dust clouds, and the nature of dust grains. Spectra of
luminous OB stars penetrating foreground SF regions provide
important information on spatial distribution and physical scales of
dust in different regions. This LAMOST legacy survey, along with
various data sets in the infrared (e.g. 2MASS, UKIDSS, Spitzer,
Herschel) and sub-mm (e.g. JCMT, ALMA), provides important
information on the 3D distribution of extinction across the disk.
Once available, the 3D extinction map will allow for the first time
detailed quantitative accounts of photometric corrections along any
line of sight.

Other less luminous, but more numerous, stars in the disk can be used
to trace the three dimensional distribution of dust in the solar
neighborhood.  For brighter stars with high quality photometry, and spectra at both
R=1800 and R=5000, we will be able to disentangle the star's temperature,
surface gravity, [Fe/H], [$\alpha$/H], and reddening.  The reddening along
each line of sight will be an important tool for mapping the local dust
extinction.

\subsection{A complete census of comparatively evolved YSOs }

The Herbig \& Bell Catalogue (HBC; Herbig \& Bell 1988), which has long been serving
as an important source for detailed investigations of the formation
and early evolution of solar-like stars, included 735 emission line stars
serendipitously discovered toward the Orion constellation.
LEGUE could produce a nearly complete census of
comparatively evolved YSOs, classical T Tauri stars (CTTS) and weak
lined T Tauri stars (WTTS), over the portion of the Galactic plane
surveyed.  This data will be
a treasury legacy of LAMOST for studies of the history of
large-scale star and cluster formation across the disk. The spatial
distribution of WTTS, combined with results from multi-wavelength
studies of star formation in e.g. the infrared and/or sub-mm
observations will provide important clues to our understanding of how
star and cluster formation propagates through giant molecular
complexes.

\section{Observing Constraints}

The LEGUE survey is designed to serve the science objectives, subject
to the constraints of the telescope system (Zhao et al.
2012).  Since the telescope is constrained to operate within 2
hours of the meridian, and a typical exposure set of three 30-min exposures takes about two hours to observe, there is a very small range of right ascension that can be observed at any given time.  This fact combined with the weather patterns (see Yao et al. 2012) means that the
telescope will have very little observing time towards the Galactic Center, since the Galactic center is observable only in the summer when it is almost always too humid to observe.  Most of the available clear weather will be when the
Galactic anticenter is up in the winter months.

Because Mirror B (the downward facing Schmidt primary) is fixed, the effective collecting area of the telescope and the quality of the point spread function depends on the hour angle and declination of the observations.  The largest collecting area and smallest point-spread function (PSF) are when Mirror A (the Schmidt corrector) and Mirror B are nearly aligned at $\delta=-10^\circ$.  However, at very low declinations the air mass is large, so there is high atmospheric extinction and distortion. Thus, considering all factors, the optimal image quality is to be found at around $20 < \delta < 30^\circ$.  At all declinations, $20\%$ of the light is lost at the edges of the field of view due to vignetting.

At declinations above $\delta=60^\circ$, the PSF at the edge of the focal plane becomes so large that observations are restricted to a $3^\circ$ field of view instead of the full $5^\circ$ field of view.  Because the fiber positions are nearly fixed, we lose 64\% of the fibers as well as 64\% of the field of view.  The best observing conditions are expected  in a declination range $10^\circ<\delta<50^\circ$ and within two hours of the meridian, with the additional constraint that for $10^\circ<\delta<20^\circ$ the observations are within 1.5 hours of the meridian.  Within this range, there will be a reasonable consistency of observations and a PSF that puts most of the light inside the fibers.  Outside of this range, the performance is largely unknown, and in many cases could be substantially worse than this recommended operating region.

The fibers are positioned with robotic arms that operate within 3.15' of their nominal
positions, which are about 4.7' apart.  The survey targets must be fairly evenly distributed on the sky; at most seven
fibers can be placed within any circular sky area with radius of about 3.15'.  If an open cluster is 20' in diameter, we can place at most about 50 fibers on stars
within the cluster's diameter, and those must be fairly uniformly distributed over the
cluster area.  We can select fewer than 20 targets in the vicinity of a globular cluster with
tidal radius of 10'.  In addition to limiting our ability to sample stars in clusters, the uniformity constraints of the fiber positioning system make it difficult to do completely filled surveys of any area
of the sky. 

In a single pointing, LAMOST places 4000 fibers in 20 deg$^2$ of sky, which comes to about 200 fibers deg$^{-2}$. This is comparable to the fiber density in SEGUE, though SEGUE required two visits to each position on the sky to achieve 180 fibers deg$^{-2}$ (note also that with multiple visits, LAMOST can easily increase the target density to 600 deg$^{-2}$ or more).


The spectrographs are designed for $R=1000$ gratings and $R=5000$ gratings.  The wavelength range is 3700 to $\sim9000$ \AA~for $R=1000$ gratings.  Because we need better resolution than this to achieve the required velocity resolution and to measure elemental abundances, we place slits that are 2/3 of the fiber diameter in front of the fibers. In theory, this should block 20\% of the light and  corresponds to the resolution of $R=1800$.  The LEGUE spectra will be similar to SEGUE spectra, with radial velocities and
metallicities determined to $\sim$ 7 km s$^{-1}$ and 0.3 dex, respectively.

The $R=5000$ gratings will yield two pieces
of the spectrum that are $350$~\AA~wide, one in the red and one in
the blue.  The blue wavelength coverage is centered around
$5300$~\AA~to sample the metal lines, including Mgb (5175~\AA).  The red
segment covers the spectral range 8400-8750~\AA, sampling the
CaII (triplet), FeI, TiI, etc., which are ideal for measuring the RV, $\mathrm{[Fe/H]}$, and detailed chemical abundances.  The red portion of the $R=5000$ wavelength coverage and resolution are comparable to that of the RAVE
experiment.  The accuracy in measuring RV and $\mathrm{[Fe/H]}$ at $R=5000$
are expected to be 1 km s$^{-1}$ and 0.1 dex, respectively.
These gratings will not be available in the first year of the survey, so $R=5000$ observations will not start immediately.

Due to weather constraints, we have very little opportunity to
observe at right ascensions of 16h to 22h.  The part of the sky for which we will have the
greatest number of observations is 2h to 8h, when fields near the Galactic anticenter are
available. When planning the survey,
we must ensure that there are always fields available at each right ascension when it becomes
available.


\section{Spheroid, Disk, and Anticenter Components}

The LEGUE spectroscopic survey is divided into three major parts, which have different magnitude, selection, and signal-to-noise requirements.  The three components of the Galaxy that will be surveyed by LEGUE are (1) the spheroid, (2) the Galactic anticenter, and (3) the disk.  The spheroid science requires the faintest targets, and is most similar to the SEGUE survey.  The anticenter survey takes advantage of the fact that the majority of the good weather at Xinglong is in the winter months when the Galactic anticenter is high in the sky to do a statistical sampling of this important part of the Galaxy.  The disk survey will sample bright stars, particularly those in open clusters, when the moon is bright.  In the disk there are enough stars to populate the LAMOST fibers even at bright magnitude limits.  In each of the three components, we will strive to achieve the best possible science with every fiber and photon that is available to the telescope while targeting stars of all types to provide a large serendipitous discovery space.

\subsection {Spheroid Survey}

We will perform a spectroscopic survey of at least 2.5 million  stars selected from SDSS (I,II,III) imaging with $|b|>20^\circ$, at a density of 320 stars per square degree or higher, over two contiguous areas: one in the north Galactic cap and one in the south Galactic cap, totaling 5000 square degrees of sky or more.  In the north Galactic cap, brighter stars will be observed when the weather is not pristine and the footprints of the disk/anticenter surveys are not visible in the sky.  At least 90\% of the survey plates will be in this contiguous area, and at least 90\% of the science fibers on each plate will be assigned based on a set of uniform survey criteria,
using $r$, $(g-r)$, and $(r-i)$. Using these criteria, we can target essentially all blue O, B, A, and WD stars, and a statistically significant fraction of the high latitude F turnoff, K giant, M giant, and $0.1 < (g-r) < 1.0$ stars.

Because we cannot observe all possible targets, we will employ weighted random sampling to select stars from all possible spectral types and classes (see Carlin et al. 2012).  There will not be separate individual selections for each type of star, as was used in SEGUE I.  The target selection algorithm will be similar to the spheroid target selection algorithms used for the pilot survey (Zhang et al. 2012, Yang et al. 2012).  It will still be complicated to calculate the fraction of spectra observed for any given color, magnitude, and position on the sky, because the fraction of stars observed will depend on the star density and the number of times that part of the sky was observed; in higher density regions and the first plates observed in a given part of the sky, a higher fraction of the spectra will be of relatively rare objects in less populated regions of color and magnitude.

\subsection {Anticenter Survey}

The anticenter survey will cover the region $150^\circ < l < 210^\circ$ and $-30^\circ < b < 30^\circ$, sampling a significant volume of the thin/thick disks as well as the halo. We will use the Xuyi photometric survey to select target stars, aiming for an even coverage across multidimensional $r, g-r, r-i$ color-magnitude space as well as in spatial distribution on the sky whenever possible, to minimize selection biases.
We plan to survey an average of 1000 stars per square degree for $|b| > 2.5^\circ$ and twice that for lower Galactic latitudes. In total, approximately 3.7 million stars in the ~3600 square degree region will be surveyed, of which about a third (1.2 million) are faint and 2.5 million are bright.  There is significant overlap between the Disk Survey and the bright portion of the Anticenter Survey.  There is also significant overlap between the faint portion of anticenter survey and the spheroid survey.

\subsection {Disk Survey}

We define the disk survey as the Low Galactic Latitude Survey,
which will observe as much of the disk with $-20^\circ < b <
20^\circ$ as can be covered from Xinglong, given the latitude and
weather constraints, and making sure to include all known open
clusters in this region (see Chen et al. 2012 for an overview of the disk survey). Therefore the region for $150^\circ \le l
\le 210^\circ$ will be a subset of the Galactic anticenter survey,
for which the input catalog will be selected from the Xuyi
photometric survey. If this survey is not available in a particular
region of the sky, then targets will be selected from UCAC3 and
2MASS. Note that from Xinglong station, the $20^\circ \le l \le
80^\circ$ region of the disk will be poorly sampled, due to a
limited number of clear nights in summer.

We will perform a spectroscopic survey of at least 3 million $r<16$
Galactic disk stars. The total region of sky available to survey is
about 6000 square degrees. We would like 1000 objects per square
degree (requiring five visits to each field), which means we will
likely only cover 3000 of those square degrees. Weather will dictate
that the area near the anticenter will get good coverage. The
priority for the other regions will be to get lower latitude stars
because that is where the open clusters live. The highest priority
targets will be open cluster members; the remainder of the fibers
will be placed on stars using a weighted random sampling of optical
color and proper motion.



In addition, we will obtain $R=5000$ spectra of stars that have
already been observed at lower resolution, starting in year two or
three of the survey.  By that time, we will have multiple coverings,
so the most interesting objects from the $R=1800$ survey can be
selected.  The $R=5000$ data are most useful if we already have $R=1800$
data, since $R=5000$ will give precise alpha and other elemental
abundances, once the other stellar parameters are known from
lower-resolution spectra.  The photometry is not sufficient to get
temperature in highly reddened regions.   However, if we have $R=1800$
and $R=5000$ spectra combined with XuYi photometry, then we will be
able to determine accurate stellar parameters, alpha abundances, and
reddening for each star. Furthermore, the study of internal open
cluster kinematics requires 1-2 km/s radial velocities (on the order
of the velocity dispersions in open clusters) provided by $R=5000$
spectra.

Since each fiber has a limited range of motion in the focal plane,
we need four plates on each open cluster to obtain enough data to
study their properties.  These clusters are extremely important for
calibration of the radial velocities as well as the stellar
parameters, and are an interesting science project in their own
right.  There are enough open clusters near the Galactic plane that
there is little difference between specially targeting the open
clusters and surveying all areas of the sky.

By necessity, the majority of the disk fibers will be placed on
stars that are not in open clusters.  These spectra will be vital
for studying the local dynamical and chemical structure of the disk,
looking for substructure and gradients in disk properties, studying
disk moving groups, obtaining a complete sample of young stellar
objects, and studying disk dust and extinction in three dimensions.

\section{Sample Survey strategy}

\begin{figure}
\centerline{\includegraphics[width=12cm]{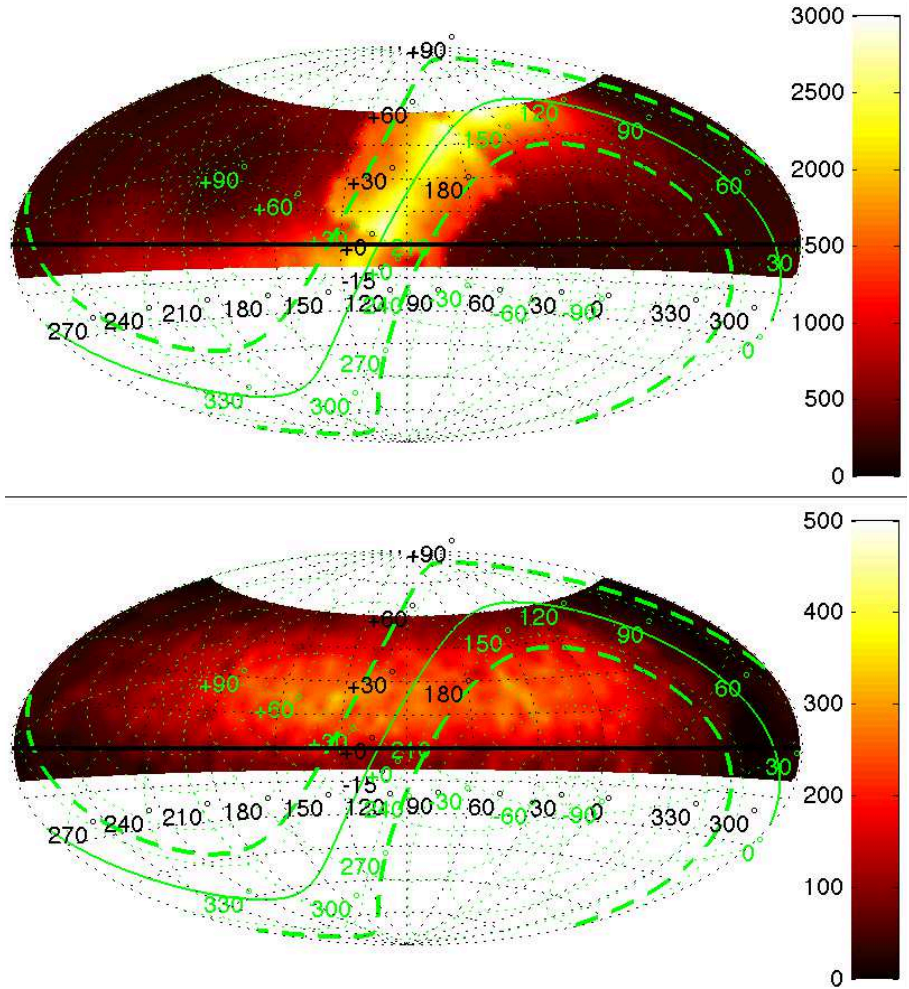}}
\caption{The expected number density of LEGUE survey spectra (spectra/deg$^2$) in the whole visible sky with
weather conditions and moon phase
considered (see Yao et al 2012, this volume). The density is coded in color, with the scale shown in the sidebar. The Galactic plane is shown by the green solid thin line. The two
dotted green lines indicate the limit between halo and disk regions of the survey (set at b=$|20|^\circ$).  The upper panel shows the predicted number density of spectra observed on bright nights, when the moon is above the horizon and whenever the sky is not perfectly transparent; resolutions at both R=1800 and R=5000 will be used. The lower panel shows the predicted number density of spectra observed on dark nights (with R=1800 only). The high density area (in yellow) prominent in the upper panel is the Galactic anti-center area. Both panels are in equatorial coordinate; the Galactic coordinate tick marks are also given.
\label{estimate}}
\end{figure}

The number of spectra obtained is set by the number of clear nights per year at
Xinglong Observing Station, the number of fibers, the estimated exposure time, and
a reasonable length to the proposed experiment.  From a study of actual observatory
weather information over a four year period, we expect about 1700 hours or 212 nights (1400 hours or 175 nights from September through April)
per year.  About 25\% of the nights are good nights in which the faint plates can be observed.  Three quarters of the time is dark/grey.  Therefore there are about 263 hours (33 nights per year) in which we can take dark plates.  We will assume that 10\% of the good nights will be lost due to mechanical failure.

It takes one hour to adjust the mirrors at the beginning of the night in twilight, and another hour to adjust the mirrors in the middle of the night.  The overhead for readout, slew, and fiber positioning is about half an hour.  For 1.5 hour exposure times, (split into three 30 minute exposures) taken in dark/grey time, we
can obtain 3 or 4 fields per night.  The required exposure time on bright plates varies, but is typically 30 minutes of exposure for $R=1800$ and 60 minutes of exposure for $R=5000$. With this exposure time we can take 6-7 fields per clear night at $R=1800$ and 4-5 fields per clear night at $R=5000$.  After the first year, the bright targets will be re-observed with the $R=5000$ gratings in place.  The bright plates are observed when the moon is bright or when the transparency is poor.  The bright targets are observed about 80\% of the time.  With 3500 science fibers assigned per plate (the other fibers are used for calibration), we can observe 1.2 million bright targets and 675,000 faint targets per year, where we have included both $R=1800$ and $R=5000$ observations for the bright targets.

To be clear, the low and medium resolution spectra will not be taken at the same time, or even in the same year.  Changing the gratings requires that the spectrographs be re-calibrated, which will be done during daylight hours, and the $R=5000$ gratings will not be available in the first year of the survey.  In the first year of the survey, we expect to obtain 2.4 million $R=1800$ spectra of bright targets.

Figure~\ref{estimate} shows a simulation of the number of targets per $\mathrm{deg}^2$
that could be obtained, with these selection criteria over the course of a five year
LAMOST survey, taking weather and moon statistics into consideration,
in Equatorial coordinates (Aitoff projection).
The simulation was made for the 5 years 2012.10-2017.10. The constraints due to weather and other site conditions used for the simulation are discussed in detail by Yao et al. (2012, this volume).

The assumptions used to build the model in Figure~\ref{estimate} were more optimistic than we used in this proposal.  All of the observations in the Galactic plane assume 10 plates observed
per clear, moonlit night, and the observations in the spheroid assumed that all clear weather had high transparency and 4 plates were observed per night.  The large number of fibers in the anticenter region reflects
our priority for observing a statistical sample in this region.

In the LEGUE survey, we will make the survey area as contiguous as possible.  At any given time, the right ascension of the observations is nearly fixed, but there is freedom to change the declination.  We will begin our observations at a fixed declination.  As more plates are observed, we will add on to the contiguous area to the north and south.  The low latitude regions of the survey (which are near the anticenter) will be covered many times, so the survey strategy will optimize contiguous and uniform coverage.

The survey time may be shared with an extragalactic survey, though the start of the extragalactic survey is likely to be delayed.  The extragalactic survey (LEGAS) will operate only in the spheroid region, and could take as much as 2/3 of the dark/grey observing time once it begins.  
The LEGAS survey does not have high enough target density to use all of the fibers in each plate, so remaining fibers are allocated to LEGUE stellar targets. In practice LEGAS is unlikely to use more than about half of the dark/grey fibers.

\section{Future Planning}

The LEGUE survey will begin in fall 2012.  The survey will include at least 2.5 million fainter spectra in the spheroid and anticenter regions, and at least 5 million brighter spectra, concentrated towards the Galactic plane.  The target selection algorithms will be updated versions of the target selection algorithms used for the pilot survey, informed by the pilot survey results.  Due to the uncertainty in the start date of the extragalactic (LEGAS) portion of the survey, the length of time required to complete the survey is uncertain.  At the rates estimated in this paper, the survey can be completed in four years, but the actual time to completion may be longer if LEGUE does not use all of the telescope time.

\acknowledgement
We thank the guest editor of this mini-volume and the referee of this paper, Joss Bland-Hawthorn for helpful comments on the manuscript.
This work is partially supported by Chinese National Natural Science
Foundation (NSFC) through grant No.10573022, 10973015,11061120454, 11243003, and the US National Science Foundation, through
grant AST-09-37523. 



\end{document}